\documentclass[12pt]{iopart}

\usepackage{iopams}  
\usepackage{graphicx}
\usepackage{dcolumn}
\usepackage{bm}
\usepackage{hyperref}
\usepackage{color}
\usepackage{subfig}

\begin{document}
\title[Knots in Steady Shear Flows]{Dynamics and Topology of Flexible Chains: Knots in Steady Shear Flows}

\author{Steve Kuei$^{1,2}$, Agnieszka M. S\l owicka$^3$, Maria L. Ekiel-Je\.zewska$^3$, Eligiusz Wajnryb$^3$ and Howard A. Stone$^1$}
 \address{$^1$ Department of Mechanical and Aerospace Engineering, Princeton University, Princeton NJ 08544}
 \address{$^2$ Rice University, Department of Chemical and Biomolecular Engineering, Houston, TX 77005}
\address{$^3$ Institute of Fundamental Technological Research, Polish Academy of Sciences, Pawi\'nskiego 5b, 02-106 Warsaw, Poland}
\ead{mekiel@ippt.pan.pl, hastone@princeton.edu}


\begin{abstract}
We use numerical simulations of a bead-spring model chain to investigate the evolution of the conformation of long and flexible elastic fibers in a steady shear flow. In particular, for rather open initial configurations, and by varying a dimensionless elastic parameter, we identify two distinct conformational modes with different final size, shape, and orientation. 
Through further analysis we identify slipknots in the chain. Finally, we 
provide examples of initial configurations of an ``open" trefoil knot that the flow unknots and then knots again, sometimes repeating several times. These changes in topology should be reflected in changes in bulk rheological and/or transport properties.

\end{abstract}


\vspace{2pc}

\section{\label{Intro}Introduction}

One area of complex fluids concerns the motion and topology of elastic filaments in fluid flows, which is inspired by both natural and industrial phenomena, such as flagellar motion \cite{CamaletJulicherProst,WolgemuthPowersGoldstein,SpagnolieLauga2010}, fluid-structure interactions, e.g. \cite{ShelleyZhang,LindnerDuRoure}, and polymer processing, e.g. 
\cite{MaGraham,Reddig,Ladd2006,somani_flow-induced_2005,zhang_numerical_2011}. Such long flexible filaments can have non-trivial dynamics, e.g. drift across streamlines can occur, e.g. \cite{Reddig,JendrejackGraham,slowicka_lateral_2013}, and topologies. For example, knots are found in Brownian systems such as bacterial DNA, protein structures, and polymer chains, and non-Brownian systems such as ordinary string, elastic  fibers and chains of linked beads \cite{BelmonteShelleyEldakarWiggins,HickfordJonesEggers}. In this paper we report the time-dependent shapes of flexible non-Brownian filaments in a steady shear flow and identify conditions that allow an ``unknotting-knotting" transition, where an unknotted filament is later observed to form an (open) knot as it rearranges continually in the flow.

For rigid non-Brownian particles in shear flow, when the Reynolds number is low, a straight elongated particle has a Jeffery orbit, which is the basis for many studies, and helps organize those initial states that have orientation and rolling about the vorticity axis from those that primarily align with and tumble periodically about the flow direction. For flexible or deformable objects, characterizing the shape of the particle in flow, and the corresponding dynamics, is more challenging. For example, experimental measurements of fiber dynamics commonly use DNA as a model polymer \cite{shaqfeh_dynamics_2005}, and,  on a larger length scale where Brownian effects are less significant, linked colloidal chains are also used as experimental models of a worm-like chain \cite{biswal_mechanics_2003, li_bending_2010} and the cellular flagellum \cite{Dreyfus}. In all of these cases,  dynamics in a single plane are measured, but it is difficult experimentally to simultaneously track dynamics in both the shear and vorticity directions \cite{teixeira_shear_2005}.

On the other hand, numerical simulations make possible the visualization of multiple viewing planes and the detailed analyses of shape and topology of flexible filaments. Previous work, some of which includes Brownian effects and some of which does not, has identified and tracked the tumbling of flexible filaments (e.g. polymers) in detail, relating the tumbling period and shape to polymer length, flexibility and shear rate, as well as rates of rotation both in and out of plane \cite{slowicka_lateral_2013,liu_brownian_2004,matthews_complex_2010, huang_non-equilibrium_2012,slowicka_dynamics_2012}. However, due to computational constraints, these simulations of fibers are generally bead-chains with about $N = 50$ beads, with the longest ones containing $N= 100$ beads; an alternative approach to some of these questions is to use slender-body theory, e.g. \cite{BeckerShelley,YoungShelley}.  

In recent years, knots in long chains has been studied more widely, including the observation of thermally driven knotting and unknotting of a polymer chain \cite{TubianaRosaFragiacomoMicheletti2013}, the influence of chain tension on the knot, e.g. \cite{MatthewsLouisYeomans,MatthewsLouisLikos}, and the influence of electric fields on knots \cite{TangDuDoyle2011}. Also, the untying of a knot in an extensional flow has been studied recently \cite{RennerDoyle}, and has the spirit of the present paper where we ask about the effects of flow. In contrast, for non-Brownian systems other kinds of forcing, such as vibration, can produce knotting of chains of linked spheres \cite{BelmonteShelleyEldakarWiggins,HickfordJonesEggers}. We are not aware of any similar studies, experimental or numerical simulations, highlighting conditions that lead to long non-Brownian chains forming knots in flows.

Thus, while significant steps have been taken towards understanding the detailed dynamics of flexible non-Brownian fibers, to understand topology is much more difficult: it requires three-dimensional detail and very flexible, long and thin fibers as, in theory (when there is no flow), the probability of knotting in a random walk increases exponentially with the length of the walk \cite{Delbruck1962}. Moreover, a smart choice of initial conditions is needed and an operational definition is needed for a knot in an open fiber. 

In this paper, we focus on the dynamics and topology of long, flexible, non-Brownian chains of beads in a steady shear flow and use numerical simulations to investigate whether knots in the fiber occur as a result of the flow. We will find that unknotted fibers are capable of forming (open) knots, and in some cases we document a sequence of unknotting-knotting transitions. The numerical solution of this kind of problem requires tracking the motion of $N$ beads, where hydrodynamic interactions between the beads mean that the dynamics  of the topology of the object involve $3N$ coupled nonlinear ordinary differential equations. Not surprisingly, such dynamics should be expected to be chaotic though we have not attempted in this paper to link our study of complex shapes and the unknotting-knotting transition to the underlying chaotic dynamics.  Our results highlight new aspects of the dynamics of flexible filaments in flow.  

\section{Simulation methods}

\subsection{A fiber as a string of spherical beads}

\begin{figure}
\subfloat[]{\label{coorda}\includegraphics[width=0.5\textwidth]{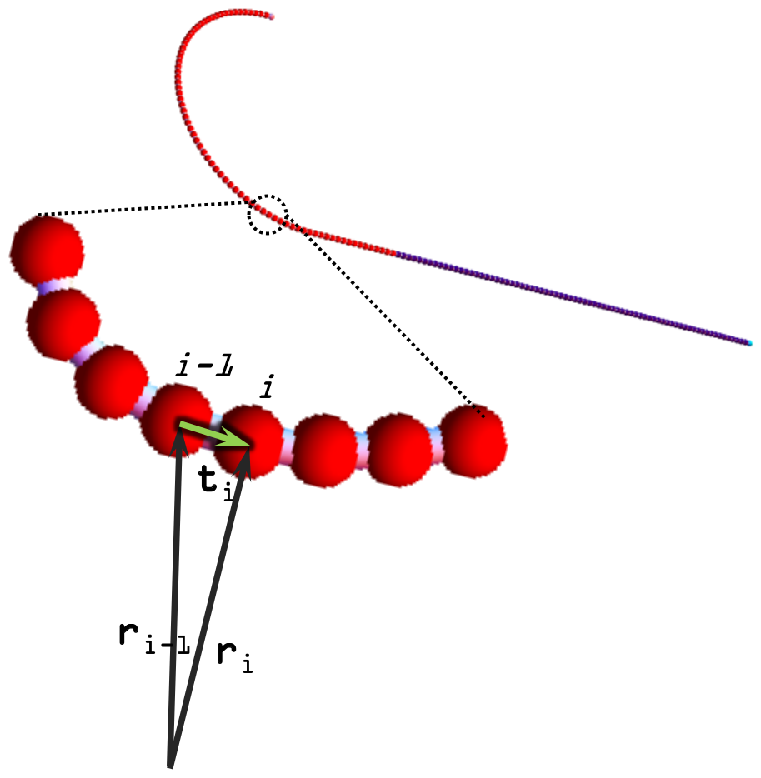}}
\subfloat[]{\label{coordb}\includegraphics[width=0.5\textwidth]{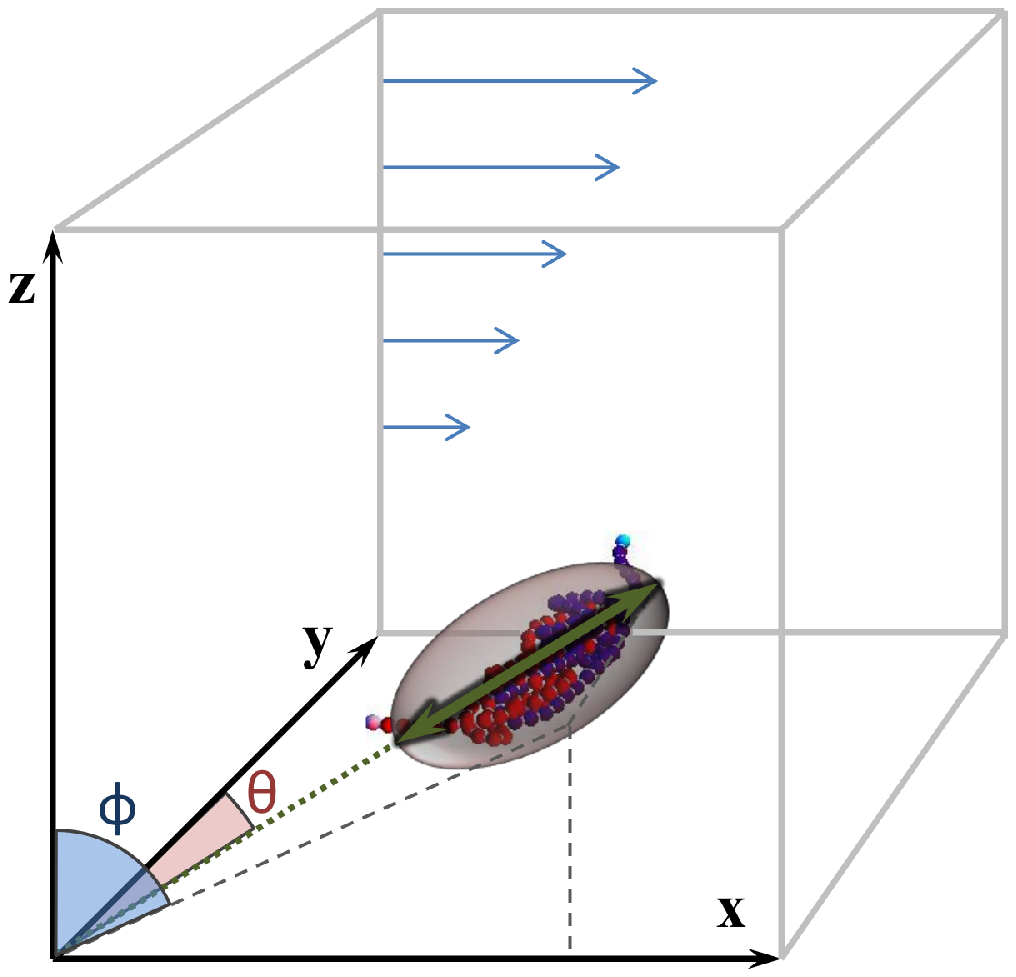}}
\vspace{-0.2cm}
\caption{Coordinate systems and notation for the simulations. (a) A  `candy-cane' shape, which is one of two typical initial configurations studied in this work. The expanded view labels beads $i$ and $i-1$. Also, $\mathbf{r}_i$ is the position vector of each bead center and $\mathbf{t}_i$ is the vector connecting two adjacent bead centers. (b) A configuration with shading to indicate the chain's ellipsoid of revolution. The shear flow is shown in the background. The major axis of the chain's configuration is shown with an arrow, while the red and blue coloring of beads denote two halves of the fiber for visual clarity. The orientation is measured by the angles $\theta$ and $\phi$, such that $\theta$ measures the inclination from the vorticity axis ($y$) and $\phi$ measures the orientation in the $xz$-plane, measured from the $z$-axis.}
\label{coords}
\end{figure}

We consider a long non-Brownian flexible fiber in steady undisturbed shear flow ${\bf v}_0 =\dot{\gamma} z{\bf e}_x$ of a fluid with viscosity $\eta$ (figure \ref{coords}). The vorticity vector is in the $y$-direction. The fiber deforms owing to elastic and bending forces and is  modeled as a chain of $N$ identical spherical beads with diameter $d$ (figure \ref{coords}a) \cite{slowicka_lateral_2013,stark2006}. The parameters of these forces are $k$, the spring constant normalized by $\pi \eta d\dot{\gamma}$, $A$, the bending stiffness normalized by $\pi \eta d^4 \dot{\gamma}$, and $\ell_0$, the equilibrium distance between the closest bead centers measured in units of $d$.  All distances are measured in units of $d$, and time in units of $1/\dot{\gamma}$. Thus, there are four dimensionless parameters that characterize the chain: $k$, $A$, $N$ and $\ell_0$. We work close to the inextensible limit with $\left |\ell_0-1\right | \ll 1$ and $k\gg 1$ and focus on the dynamics for a wide range of values of $A$.

We assume that the particle-scale Reynolds number is much smaller than unity, i.e. $Re=\frac{\rho {\dot\gamma}d^2}{\eta}\ll 1$, where $\rho $ is the fluid density and the fluid flow satisfies the quasi-steady Stokes equations, with no-slip boundary conditions at the bead surfaces. The equations are solved using the multipole expansion method, which accounts for lubrication corrections to speed up the convergence of the friction matrix as the order of the multipole truncation is increased. This solution yields the velocities of the beads, as explained in \cite{slowicka_lateral_2013}. The algorithm and its numerical implementation {\sc hydromultipole} are described in \cite{cichocki_lubrication_1999,fs}. Also, the application of these tools to model dynamics of flexible fibers is outlined in \cite{slowicka_lateral_2013}. The time-dependent translational velocities of the bead centers, evaluated by the {\sc hydromultipole} numerical code, are used by an adaptive fourth-order Runge-Kutta procedure, from which the positions of every bead are updated. 

We focus on the dynamics of long fibers ($N \!\approx \!100\!-\!150$), which, in flow, tend to fluctuate from elongated shapes to compact shapes with some of the beads coming  close together. Spurious overlaps of beads can occur, which can significantly influence the time scales of the dynamics of close particles~\cite{ekiel-jezewska_stokesian_2008}. Therefore, we analyzed different procedures that allow the dynamics to be continued once an overlap occurs and chose a procedure that caused the least number of overlaps. If the distance between two bead centers $r<1$, then the lubrication correction for this pair is evaluated for the rescaled configuration $1+\epsilon_{lub}$, where $\epsilon_{lub}=10^{-4}$ is the integration accuracy per step, and the dynamics of the original positions are continued. The details of this original simulation approach for handling close particle pairs are provided in the next section. 

In the simulations reported here,  the initial geometry is chosen first from a random rotation of 
a `candy-cane' shape with $N=152$, i.e. a straight chain that bends into a semi-circle at one end; see figure \ref{coords} and \ref{AppendixCandyCane}. The candy-cane shape is convenient for starting with an elongated initial configuration that breaks the symmetries associated with shear flow. Also, we have performed calculations with a trefoil as an initial condition with $N=99$ (see Section \ref{ResultsTrefoil} and \ref{AppendixTrefoil}).
For each simulation we fix $ \ell_0 = 1.02$ and $k = 500$, and vary $A$ systematically. We evolve the shape in a steady shear flow using the {\sc hydromultipole} algorithm \cite{cichocki_lubrication_1999}. 

\subsection{A simulation procedure for compact structures}

As introduced in the previous section, and will be clear when the many simulations in this paper are shown,
long flexible fibers can form compact structures with some of the beads coming very close to other beads. As a result, it sometimes happens that surfaces of pairs of beads
 spuriously overlap. This complication will be faced by all numerical methods and for the long-time dynamics  of long flexible fibers, which are the focus of this paper, such artifacts
cannot be avoided. The physical problem is that the friction and mobility coefficients are not defined for the distance between the bead centers less than their diameter (equal to one in our units). In the literature, there exist different methods that allow the dynamics to be continued once an overlap takes place (e.g. see the brief review in \cite{ekiel-jezewska_stokesian_2008}). In this sub-section we present a new approach for the problem of overlaps; the reader interested  only in the results of the fiber dynamics, including the unknotting-knotting transition that we have identified, can skip this sub-section.

Not surprisingly, it is known that numerical procedures for handling the interaction of close pairs can sometimes significantly influence time scales of close-particle 
dynamics. In particular, examples are known of systems of particles where spurious overlaps increase the period of oscillations, even by a factor of two \cite{ekiel-jezewska_stokesian_2008}. Therefore, it is important to choose a numerical treatment of overlaps that minimizes the time-dependent number of overlapping pairs and also prevents the particle surfaces from remaining overlapped for a long time.

We have observed that spurious overlaps of beads sometimes can significantly modify compact shapes of flexible fibers, lead to their orientation along the $y$ (vorticity) axis, and change properties of 
the dynamics. In particular, a larger number of overlaps leads to a larger fiber curvature and a smaller radius of gyration. Therefore, to address the main questions of this paper we found it essential to use an optimal
procedure once an overlap takes place. After analyzing and comparing a few such methods, we proposed a new one. 
Here we provide a brief description of this procedure (called BWE), which we have applied to continue evolution of the system of equations in case of spurious overlaps of beads. Our new approach gives a  small number of overlapping pairs of the beads, which, in addition, usually do not remain overlapped for a long time. 

The procedure is based on modification of the lubrication correction ${\bm \delta}^{(2)}({\bm r}_{ij})$ in the case when the beads $i$ and $j$ overlap. To explain the idea, we first briefly explain the general idea how the lubrication correction is constructed \cite{cichocki_lubrication_1999,fs}.
Truncating the multipole expansion at the order $L$, we evaluate the multipole approximation ${\bm \zeta}^{(N)}_L$ for the friction matrix of the $N$-bead system. This step can be accomplished even if the distance $r_{ij}$ between the centers of beads $i$ and $j$   is close to but smaller than one (here, $i,j=1,...,N$). 
We need to add a lubrication correction ${\bm \delta}^{(N)}_L$ to speed up the convergence of the multipole expansion,   $\bar{\bm \zeta}^{(N)}_L={\bm \zeta}^{(N)}_L+{\bm \delta}^{(N)}_L$.  This correction is the sum of {\it all} pairwise contributions, $k\ne l$, 
\begin{equation}
{\bm \delta}^{(2)}({\bm r}_{kl})= {\bm \zeta}^{(2)}({\bm r}_{kl})-{\bm \zeta}^{(2)}_L({\bm r}_{kl}),
\end{equation}
where ${\bm r}_{kl}$ is the relative position vector of particles $k$ and $l$. Here, ${\bm \zeta}^{(2)}({\bm r}_{kl})$ and ${\bm \zeta}^{(2)}_L({\bm r}_{kl})$ are, respectively, the two-body friction matrices -- the exact one and its multipole approximation of the order $L$. 

Now, consider a pair of overlapping beads $i$ and $j$, i.e. $r_{ij}\le 1$, for which elements of the matrix ${\bm \zeta}^{(2)}$ are not defined -- they diverge at the contact. We overcome this problem by rescaling the lubrication correction for the overlapping pair of beads, 
\begin{equation}
{\bm \delta}^{(2)}({\bm r}_{ij}) \longrightarrow {\bm \delta}^{(2)}\left[(1+\epsilon_{lub})\frac{{\bm r}_{ij}}{r_{ij}}\right],
\end{equation}
and evaluate the dynamics from the unchanged positions of all of the particles. 
The multipole approximation ${\bm \zeta}^{(N)}_L$ for the friction matrix of the $N$-bead system and the lubrication corrections for non-overlapping beads are not modified.

The key idea of the method is to take $\epsilon_{lub}$ equal to  the integration accuracy per step, which in this work is equal to $10^{-4}$. Owing to this choice, the time of a spurious overlap 
is reduced in comparison to smaller values of $\epsilon_{lub}$. Indeed, a typical overlap depth is of the order 
of the accuracy. Therefore, by assigning to the relative velocity of the beads the new, quite large value, which corresponds to the gap size between their surfaces, i.e. $\epsilon_{lub}$ equal to a typical overlap depth, 
we obtain the required result that the overlapped beads can be disconnected easily, i.e.  in relative terms they quickly separate their surfaces from each other. 

\section{Results: Elongated initial configurations}

Upon starting the shear flow, the fiber, which was initially in the candy-cane configuration, described in \ref{AppendixCandyCane}, aligns approximately with the flow direction and undergoes a `tumbling' or `yo-yo' motion (figure \ref{shapesgraph}a,d; the chain halves are colored for clarity), similar to observations in studies of flexible polymers in shear flow \cite{teixeira_shear_2005,hinch1976,schroeder_dynamics_2005,gerashchenko_statistics_2006}. We identify two distinct behaviors for the dynamics, depending on the value of $A$. To quantify the dynamics, we report the time-dependent approximate shape and orientation of the chain. 

\subsection{Typical simulations and geometric characterization}

First, for a given configuration, we calculate the moment of inertia tensor \cite{rawdon_effect_2008}
\begin{equation}
T_{nm} = \frac{1}{2N^2} \sum\limits_{i=1}^N \sum\limits_{j=1}^N \left[ r_i^n - r_j^n \right] \left[r_i^m - r_j^m \right] 
\end{equation}
with the Cartesian coordinates $n,m=1,2,3$, and the labels $i, j$ of the vertices of each of the $N$ beads. The lengths $a$, $b$, and $c$ denote the three semi-axes of the ellipsoid of inertia associated with a configuration (figure \ref{coords}b). These lengths are related to the three eigenvalues $\lambda_n$ of $T_{nm}$, i.e. $a = \sqrt{3 \lambda_1}$, $b = \sqrt{3 \lambda_2}$, and $c = \sqrt{3 \lambda_3}$. The largest eigenvalue $a$ gives the primary axis of the ellipsoid, and the orientation angles of the associated eigenvector, $\theta$ and $\phi$, are measured from the vorticity and $z$ axes, respectively, as shown in figure \ref{coords}b. We use $a$, $b$, and $c$ to visualize the fiber shape: if $a \approx b \approx c$, the configuration is roughly spherical, whereas if $a \gg b \approx c$, the shape is prolate ellipsoidal, and if $a \approx b \gg c$, the shape is oblate spheroidal. 

Also, we calculate the radius of gyration $R_g$ of each configuration, i.e.  
$R_g^2 = \frac{1}{N}\sum\limits_{i=1}^N \left(\mathbf{r}_i - \mathbf{r}_{cm}\right)^2$,
by comparing the coordinates of each bead $\mathbf{r}_i$ to the center of mass $\mathbf{r}_{cm}$. Since $R_g^2={\rm trace}~ T_{nm}$ then $R_g^2 = \lambda_1+\lambda_2+\lambda_3$ \cite{rawdon_scaling_2008}. With these measures $\left\{ R_g, a, b, c, \theta, \phi \right\}$, we track a given chain's size, shape, and orientation as it evolves in the flow. 

\begin{figure*}[t]
\begin{center}
  \subfloat[]{\label{evolution1}\includegraphics[width=0.23715\textwidth]{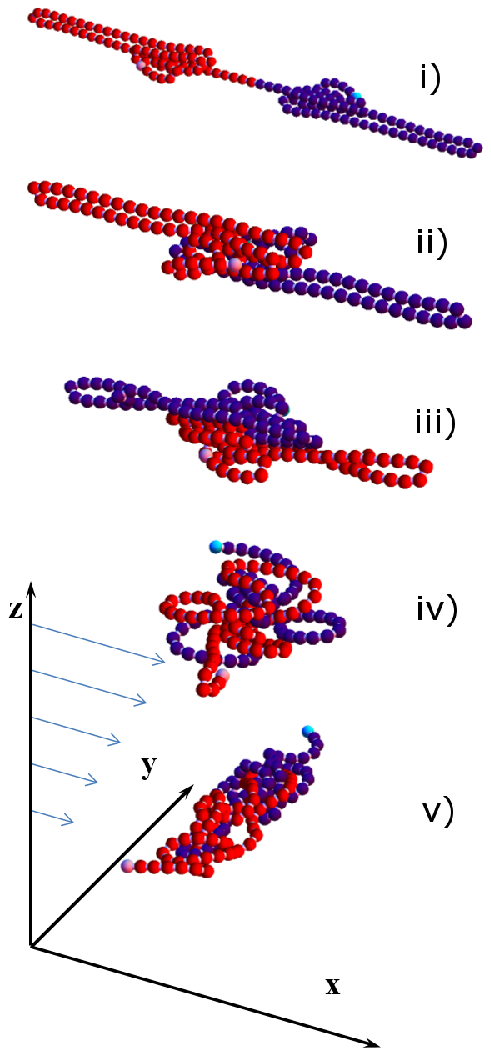}}
  \subfloat[]{\label{shapes}\includegraphics[width=0.27273\textwidth]{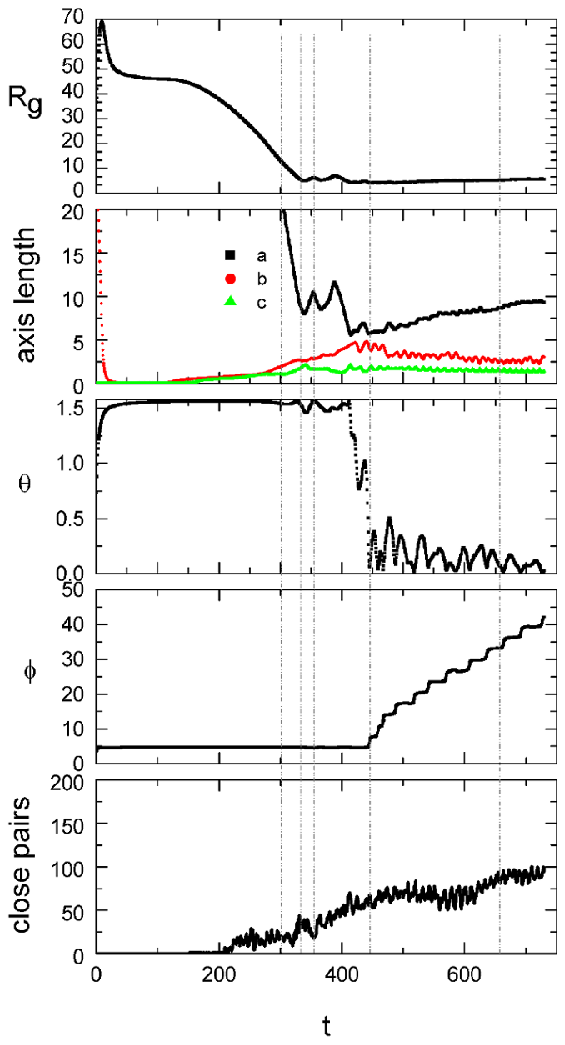}}
  \subfloat[]{\label{evolution2}\includegraphics[width=0.22925\textwidth]{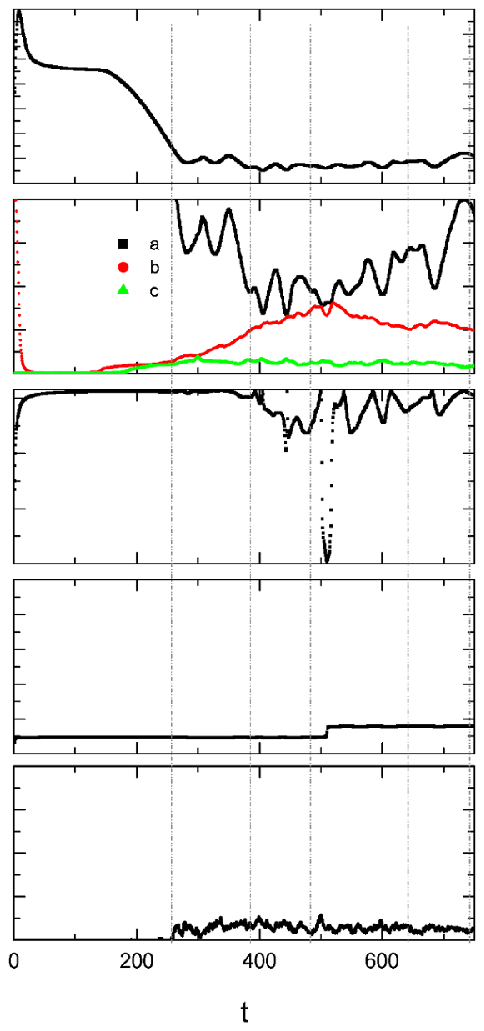}}
  \subfloat[]{\label{shapes2}\includegraphics[width=0.26087\textwidth]{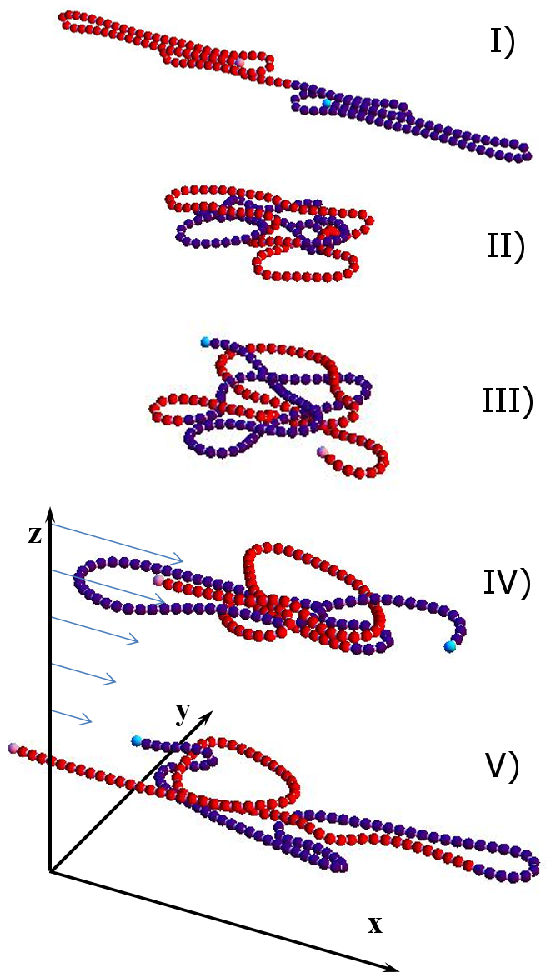}}
\end{center}
\vspace{-0.5cm}
\caption{Dynamics of chains for different values of $A$. (a) Time evolution of a chain, labeled $(i)\!-\!(v)$, with $A\!=\!1$. The first three frames show a tumbling chain aligned with the flow, as the red and blue `halves' have switched. At later times the shape is oriented along the vorticity $(y)$ axis. (b) For $A\!=\!1$, radius of gyration $R_g$, axis lengths $a$, $b$, and $c$, and the orientation of the main axis as measured by $\theta$ and $\phi$ each plotted versus time. Close pairs denote the number of beads that are within a distance $d_{ij}/d\!=\!1.05$ of another bead, not counting nearest neighbors in the chain. The dashed vertical lines denote times that correspond with the shapes $(i)\!-\!(v)$. (c) The same measures plotted as  function of time for $A\!=\!3$. (d) Time evolution of a chain, panels $(I)\!-\!(V)$, with $A\!=\!3$. }
\label{shapesgraph}
\end{figure*}

For a given initial orientation, specified in \ref{AppendixCandyCane}, chains with lower values of $A$, e.g. $A < 2$, behave differently than chains with higher values, $A \ge 3$; qualitatively, the former form tight coils aligned with and rolling in the vorticity direction, while the latter forms more open tumbling configurations, by analogy to the results of Ref.~\cite{Skjetne}. We present detailed results from each category, starting from the same initial `candy-cane'  geometry. 

First, we focus on a simulation of a more flexible chain where $A = 1$; the evolution is shown in figure \ref{evolution1} and movie \ref{evolution1} and the dynamics are quantified in figure \ref{shapes}. Initially, the chain extends and orients in the flow direction, resulting in a large radius of gyration and highly skewed axes lengths ($a \gg b, c$); the orientation angles of the main axis are $\theta \approx \pi/2$ and $\phi \approx \pi/2$, i.e. the flow direction. As time continues, the two ends fold towards the center, $R_g$  decreases significantly, while $a$ decreases and becomes comparable to $b$ and $c$. At still later times, the configuration remains relatively constant in the form of a prolate ellipsoid ($a \gg b, c$) parallel to the vorticity axis; also, $\phi$ systematically increases, which is consistent with a `rolling' motion. 

We next consider a simulation of a stiffer chain, where $A = 3$; see figure \ref{shapesgraph}c,d and movie \ref{shapes2}. The time dynamics is  different than the case of $A=1$, e.g. the collapse of the initially extended chain leads to a less compact shape that does not resemble a prolate spheroid. 
For most times, the orientation angles of the main axis remain close to the flow direction. 

Finally, we examine the number of close pairs for both $A\!=\!1$ and $A\!=\!3$. For each time-step, we check the distance between each pair of beads; we consider beads a close pair when $d_{ij}$, the distance between beads $i$ and $j$, is less than some threshold, and the beads are not the closest neighbors, i.e. $i\!\ne \!j\pm 1$. Here we set the  threshold $d_{ij}/d = 1.05$, slightly larger than the dimensionless equilibrium inter-bead spacing $\ell_0  = 1.02$. We notice that the number of close pairs in the more flexible fiber ($A=1$) is around 4 times greater than the number in the less flexible fiber (figure \ref{shapes}, \ref{evolution2}). These observations are consistent with $A=1$ yielding a tightly packed coil and $A=3$ resulting in a less compact shape.

Both visually, as seen in figure \ref{shapesgraph}a,d and movies \ref{shapesgraph}a,d, and quantitatively, through figure \ref{shapesgraph}b,c, a change in $A$ can result in considerably different dynamics. A closer inspection of the configurations shows that the chains may become self-entangled. For example, in panel (V) of figure \ref{shapes2}, the blue end has passed through a loop formed at the opposite red end. Visual inspection shows that this is a slipknot (see section \ref{SlipKonotDiscussion}), but such shapes raise the question of whether or not simple flows can cause chains to tie themselves into knots, and how to identify and classify such configurations. 

\subsection{Knot classification}

For a given configuration at a given time, we then study the topology of a chain. A knot is a topological state of a closed curve, where no set of simple line deformations, e.g. `pushing' a strand, without passing one strand through another, can change its state to that of an open circle, or an `unknot'. To classify knots, polynomials known as knot invariants, which uniquely identify each curve to a knot type, can be calculated \cite{wu_knot_1992}. Here, we calculate the Alexander polynomial of each configuration (see \ref{KnotInvariantsAppendix}). 

Any open curve, i.e. a curve with free ends, can always be `untied' by line deformations and cannot be considered knotted in a rigorous sense. However, these `open knots' still have characteristics that can be classified, and affect the properties of a chain, such as its shape, size, and relaxation modes \cite{rawdon_effect_2008}; the classification of open knots remains a subject of study. In general, an open curve must have its ends joined together to form a closed curve, from which the topological state can be calculated. There are several well-studied closure schemes \cite{lua_statistics_2006}, but as these methods can result in different topological states for the same open configuration, there is inherently ambiguity.

We use the stochastic closure scheme \cite{millett_linear_2005-1}. For each open configuration, we draw a sphere centered at the center of mass of the chain, with radius ten times the radius of gyration. A point is randomly chosen on this sphere \cite{marsaglia_choosing_1972}, and we connect by a straight line the two ends of the open curve to this point to create a closed curve that can be analyzed for its Alexander polynomial. Repeating this procedure with different random points on the sphere (here, $1000$ randomly chosen points), we build a spectrum of knots; i.e. each configuration of the flexible chain can be described as a superposition state of various knot types. 

This approach effectively provides a statistical measure of the entanglement of each chain, which we report with a knotted fraction $k_f$, i.e. the fraction of the stochastic closures which result in a knot of any type. For an open flexible fiber, ``a knot'' corresponds to $k_f$ close to $1$.  

As might be expected from visual analysis of the configurations shown in figure \ref{shapesgraph}, for the initial candy-cane configuration, no knots have been observed for the time of the simulation even though the fiber 
shapes were often very compact at long times.  The calculated values of $k_f$ were always close to zero for all time steps in the simulation.

Because the chains we study are highly flexible, it is also necessary to understand the topology of sections of the chain. Thus, 
for each subchain, the knotted fraction $k_f$ is calculated using the stochastic closure scheme and the resulting values are combined into  
a knot matrix $K(i,j)$, where $i,j = 1 \cdots N$. Each point $(i,j)$ of the matrix corresponds to starting and ending beads of the subchain, $i$ and $j$, respectively. For example, the origin, at $(1,152)$, refers to a subchain starting at $i=1$ and ending at $j=152$, i.e. the entire chain. Another point on the matrix, such as $(70,152)$, refers to a subchain starting at $i=70$ and ending at $j=152$, which is the latter half of the full chain. Therefore, every cell of the matrix $K(i,j)$ corresponds to a subchain of the original configuration. 

The knot matrix is symmetric, $K(i,j)=K(j,i)$,  because the
subchain that starts from $i$ and ends at $j$ is just the
same subchain that starts at $j$ and ends at $i$. Therefore, it is sufficient to show values of the knot matrix only for $i\le j$.

\begin{figure}[h]
\vspace{-0.3cm}
\hspace{-0.5cm} \includegraphics[width=0.5\textwidth]{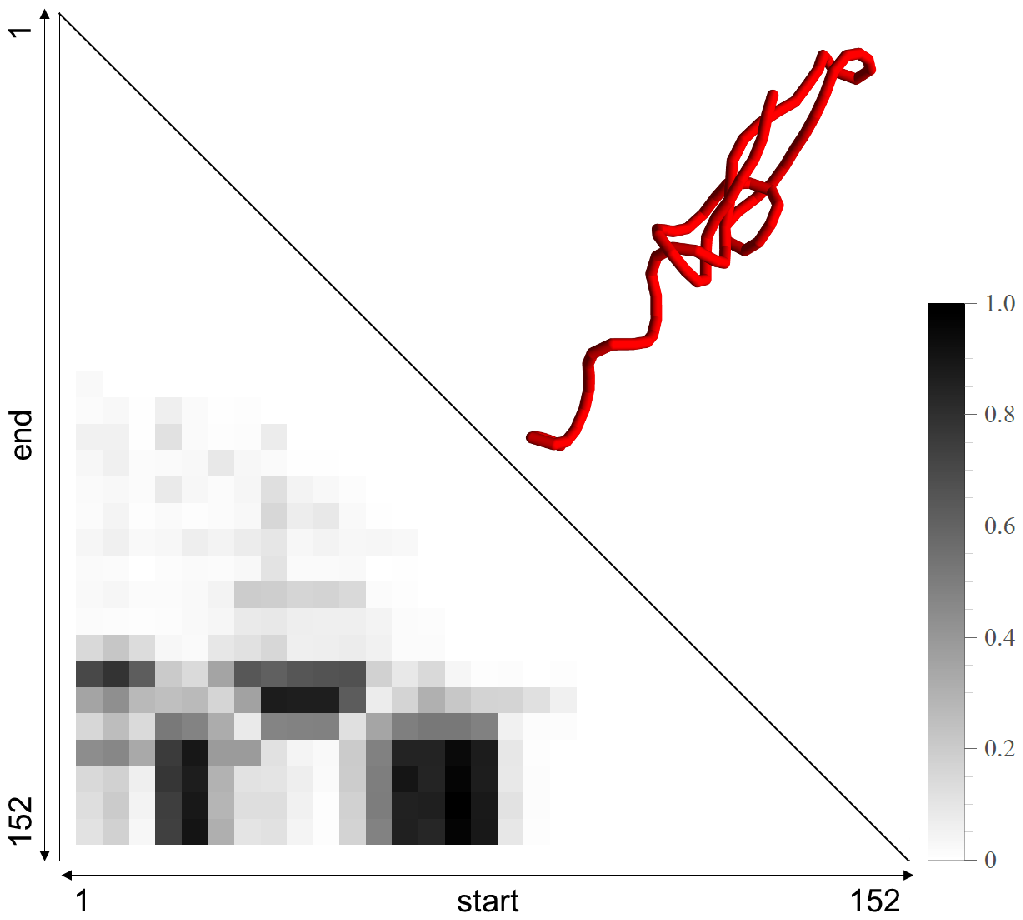}\vspace{-0.2cm}
\includegraphics[width=0.49\textwidth]{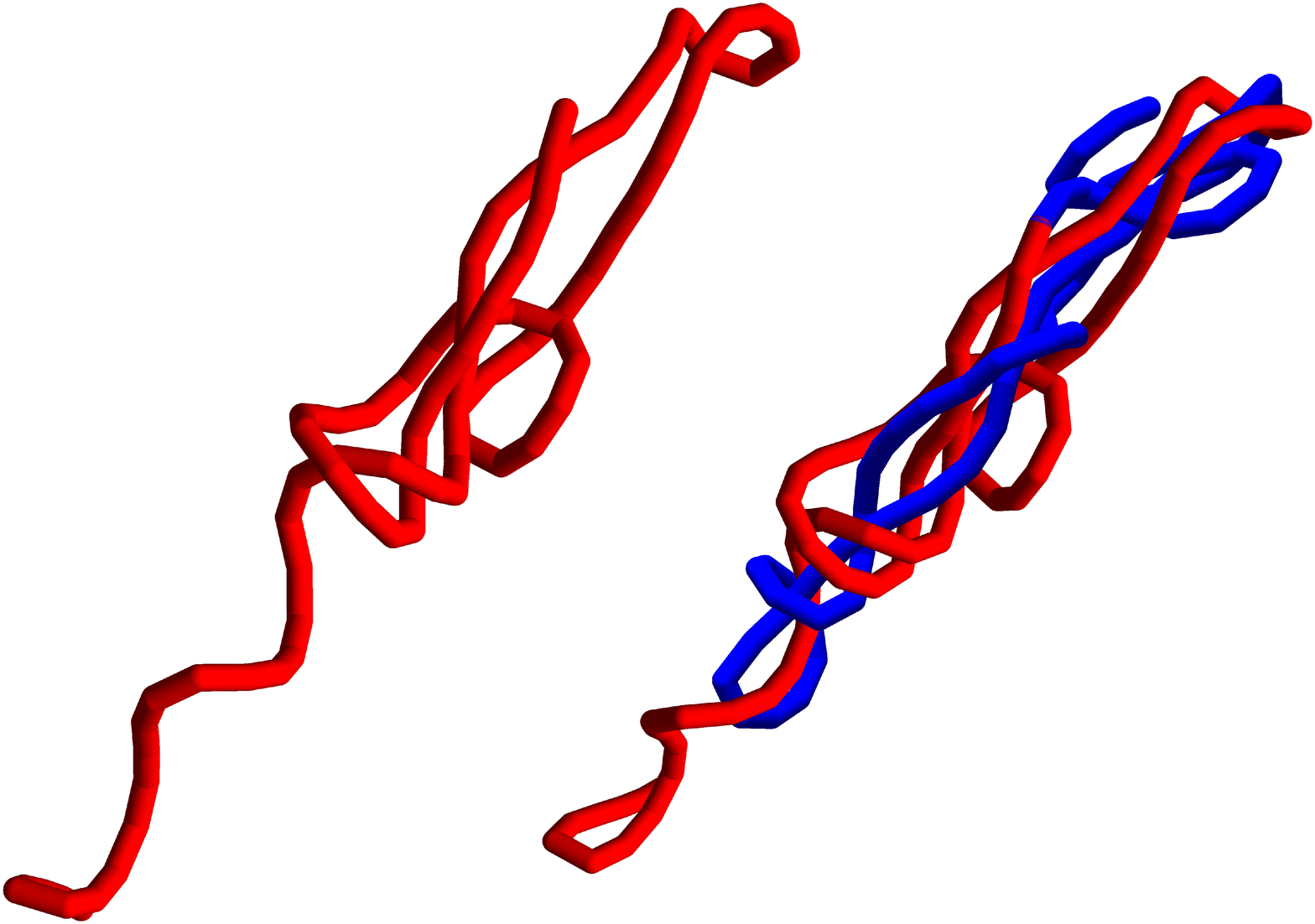}\vspace{-0.2cm}
\caption{Left: Example of a knot matrix for $A\!=\!1.4$, $N\!=\!152$, $t\!=\!682.7$. The axes denote the first and the last bead number of the subchain. The cell is colored according to its knotted fraction: $k_f=1$ is black, and $k_f=0$ is white, as denoted in the inset scale. A wire-like scheme of the knotted subchain corresponding to $(70,152)$ is shown. Right: The knot on the subchain (red) is only a slipknot in the entire chain, with the subchain pictured in red and the other sections highlighted in blue.}
\label{SlipKnotExample}
\end{figure}

\subsection{Slipknots}
\label{SlipKonotDiscussion}

In figure \ref{shapes2}(V), we observe a configuration that has the blue end of the chain threaded through a loop formed from the red half of the chain. As, intuitively, a knot requires the pulling of an `end' through a `loop', the way we might tie a knot in a shoelace, these types of configurations bear further scrutiny. 
By closing the entire chain  with the stochastic closure scheme, these configurations are shown to be unknotted - the fraction of nontrivial closures is low, $k_f$ close to $0$. 

With this visual motivation, next we studied the results of simulations at different values of $A$, keeping an eye out for possible knots.
For $A=1.4$, the knot matrix $K(i,j)$ at an intermediate time when an entangled configuration occurs is plotted in figure~\ref{SlipKnotExample}. The white cells have zero knotted fraction, while the darker cells have a higher knotted fraction. 
We note that the knot matrix is not symmetric with respect to reflection in the line $y=152-x$, because, in general,
$(i,j)$ and $(152-j+1,152-i+1)$ are different subchains, and one of them can be knotted while the other one is not.
For example, the subchains (70,152) and (1,83) are not the same and have different topological characteristics (black and white colors, respectively). 

In plots such as figure~\ref{SlipKnotExample},  dark regions can be identified as knot cores \cite{sulkowska_knotmatrix}. In these regions, the corresponding subchains have $k_f$ approaching unity and are considered knotted, even though $k_f$ is close to $0 $ for the  chain as a whole. For example, as evident in figure~\ref{SlipKnotExample}, there is a darker block corresponding to a subchain that starts at the bead 70 and ends at the bead 152. This knotted subchain is shown highlighted in red in the accompanying image in figure~\ref{SlipKnotExample}. In the right panel of figure~\ref{SlipKnotExample}, the entire chain is shown, with the knotted subchain highlighted in red, and the other part highlighted in blue.  Thus, we have identified, qualitatively and quantitatively, the formation of a slipknot in a flexible chain subjected to a steady simple shear flow. 

\section{Results: Trefoils in steady shear flow}
\label{ResultsTrefoil}

\begin{figure}[t]
\includegraphics[width=\textwidth]{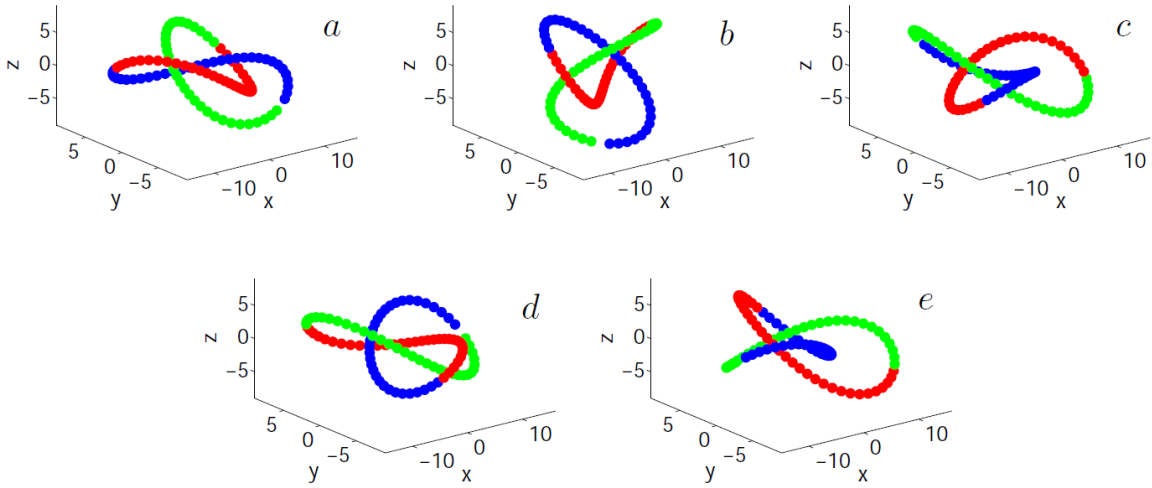}
\caption{Initial trefoil shape and five distinct orientations.}
\label{InitialConditions}
\end{figure}

Now that we have explored one extended type of initial condition, we explore a more ``knot-like" initial condition. Thus, we ask the question: can shear flow 
untie a trefoil?  We set an initial condition in a trefoil ($3_1$ knot) shape (with one bead missing),  see \ref{AppendixTrefoil}, 
and vary the initial orientation of this shape relative to the shear flow as shown in Fig. \ref{InitialConditions}. Then, we evolve the chain for 
different relative bending stiffness $A$. By using the stochastic closure scheme and counting 
invariants, we can calculate and plot the fraction knotted $k_f$ at each time step, i.e. how many of the stochastic closures result in non-trivial knots.  
A knot corresponds to $k_f$ close to $ 1$. 

We used five different orientations of this shape, shown in figure \ref{InitialConditions}a-e.  A trefoil with $N = 99$ is shown in figure \ref{InitialConditions}a and denoted as $I_0=a$.
Also, the orientations $I_0=b$ and $I_0=c$ (see figure \ref{InitialConditions}b and c) were obtained by rotating the configuration $I_0=a$ along the  $x$-axis by $\pi/2$ and $\pi$, respectively; the orientations $I_0=d$ and $I_0=e$ (see figure \ref{InitialConditions}d and e) were obtained by rotating the configuration along the  $z$-axis by $\pi/2$ and $3\pi/2$, respectively. 

\begin{table}[b]
\centering
\caption{The existence of an unknotting-knotting transition as characterized by a trefoil unknotting then knotting (1) or not (0), as a function of the stiffness $A$ and initial orientation $I_0$. The table summarizes the 29 different simulations that were performed.}
\label{ParametersShear}
\begin{tabular}{ccccccccccc}
\hline
$\;\bm{I_0 \!\!\setminus \!\!A }\!$&$\hspace{0.5cm}\bm{0.8}$&$\bm{1.0}$&$\bm{1.4}$&$\bm{2}$&$\bm{2.1}$&$\bm{2.2}$&$\bm{2.3}$&$\bm{2.4}$&$\bm{3.0}$\\
\hline
\hline
\vspace{0.1cm}
$\bm{a}$  &\hspace{0.5cm} 0  & 0 & 0 &   &   & 0 &   &   & 0 &\\
\vspace{0.1cm}
$\bm{b}$  &\hspace{0.5cm} 0  & 0 & 1 &   &   & 0 &   &   & 0 &\\
\vspace{0.1cm}
$\bm{c}$  &\hspace{0.5cm} 0  & 0 & 0 & 0 & 1 & 1 & 1 & 0 & 0 &\\
\vspace{0.1cm}
$\bm{d}$  &\hspace{0.5cm} 0  & 0 & 0 &   &   & 0 &   &   & 0 &\\
\vspace{0.1cm}
$\bm{e}$  &\hspace{0.5cm} 0  & 0 & 0 &   &   & 0 &   &   & 0 &\\
\hline
\end{tabular}
\end{table}

\subsection{Dynamics: An unknotting-knotting transition}

We performed 29 simulations starting from a trefoil shape at different orientations $I_0=a,b,c,d,e$ shown in figure \ref{InitialConditions},  
and several values of the bending stiffness $A$.  For certain $I_0$ and $A$, we observed an unknotting-knotting transition, by which we mean that the trefoil unties and then ties again. 
In Table \ref{ParametersShear}, we specify values of the bending stiffness $A$ and the initial orientation $I_0$ 
for which such a transition was (or was not) observed.  In most cases, after 
a transient period, the trefoil unties but, in some cases, at later times the chain knots again, as characterized by the knotted fraction, $k_f$, decreasing to very small values and then rising up to almost one.

\begin{figure}[t]
\includegraphics[width=1.\textwidth]{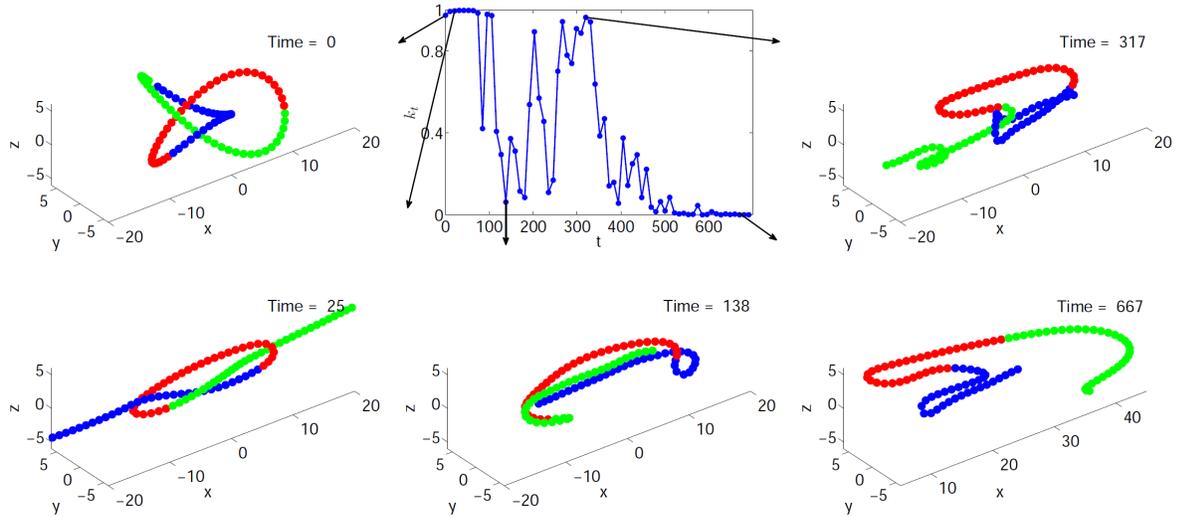}
\vspace{-0.6cm}
\caption{The unknotting-knotting transition for an initial trefoil shape, $A=2.2$, $N = 99$. The time evolution of the chain is shown, see also movie \ref{EvolutionForKnotting1}. Here $k_f$ denotes the knotted fraction based on the stochastic closure scheme described in the text.}
\label{EvolutionForKnotting1}
\end{figure}

\begin{figure}[t]
\includegraphics[width=1.\textwidth]{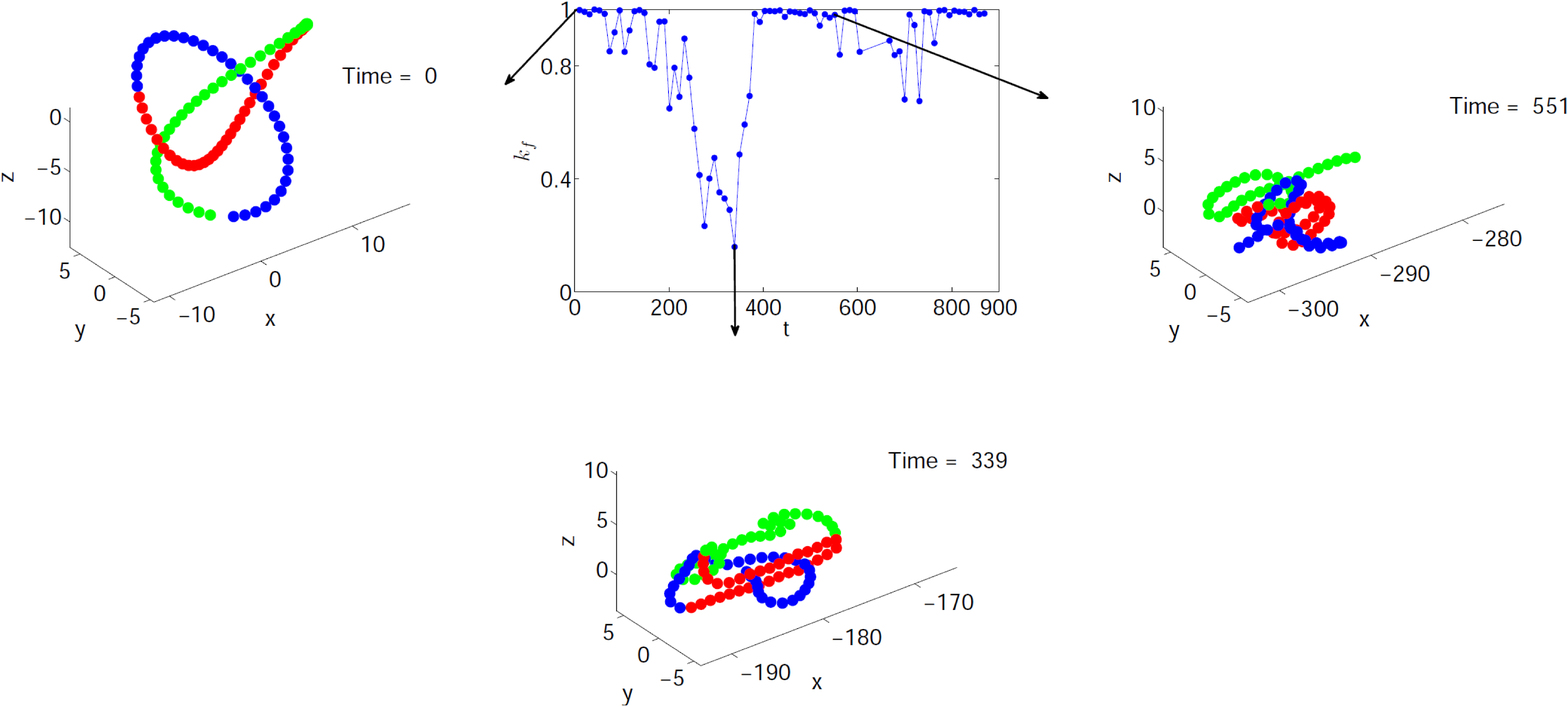}
\caption{The unknotting-knotting transition for an initial trefoil shape, $A\!=\!1.4$, $N\! =\! 99$. The time evolution of the chain is shown  (see also movie \ref{EvolutionForKnotting2}). Here $k_f$ denotes the knotted fraction based on the stochastic closure scheme described in the paper.}
\label{EvolutionForKnotting2}
\end{figure}

\begin{figure}[t]
\includegraphics[width=1.\textwidth]{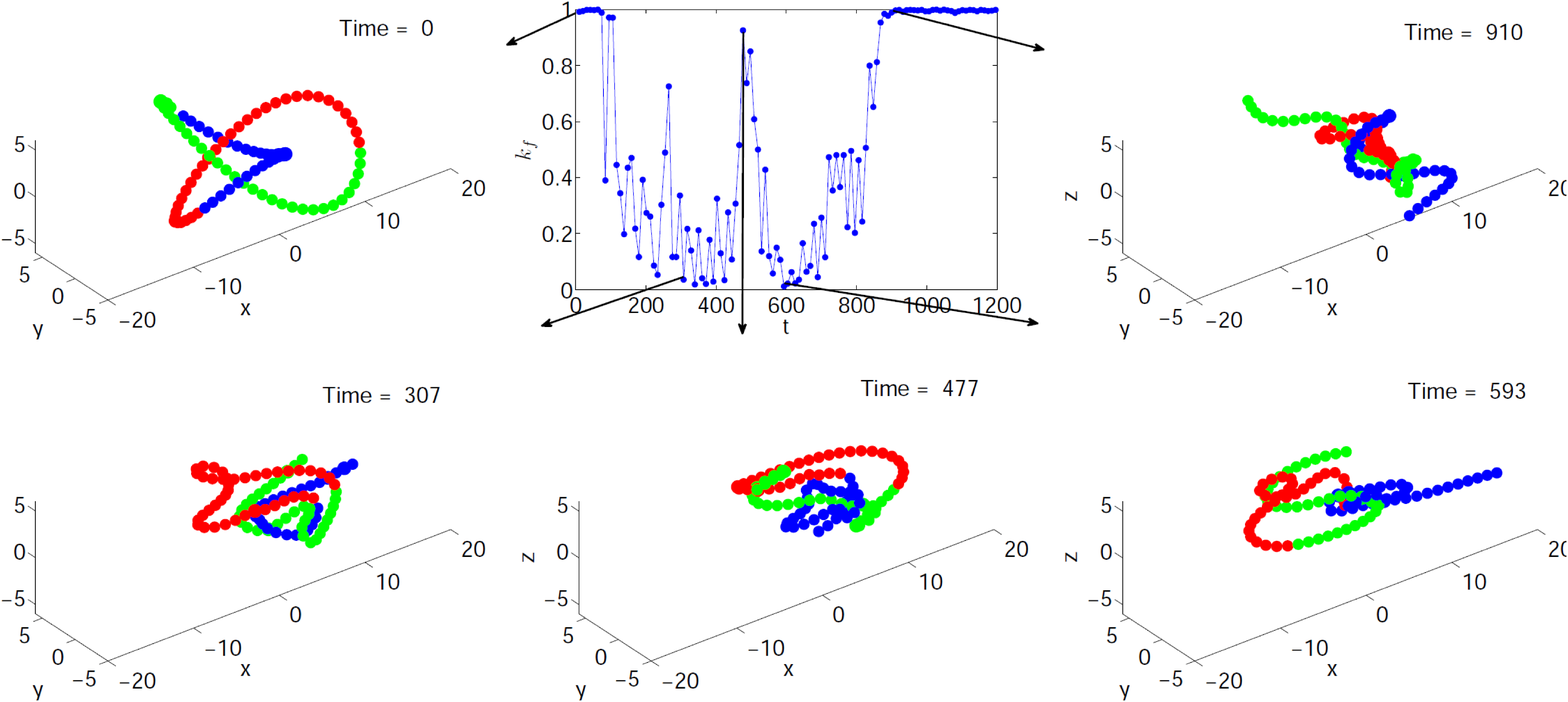}
\caption{The unknotting-knotting transition for an initial trefoil shape, $A=\!2.1\!$, $N\! = \!99$. The time evolution of the chain is shown  (see also movie \ref{EvolutionForKnotting3}). Here $k_f$ denotes the knotted fraction based on the stochastic closure scheme described in the paper.}
\label{EvolutionForKnotting3}
\end{figure}

\begin{figure}[t]
\includegraphics[width=1.\textwidth]{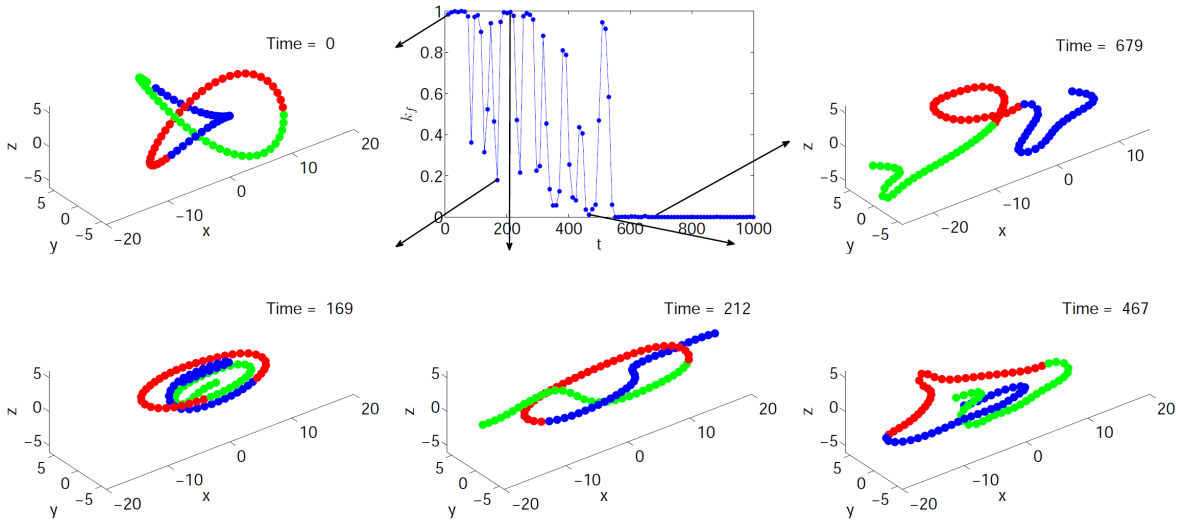}\vspace{-0.2cm}
\caption{The unknotting-knotting transition for an initial trefoil shape, $A\!=\!2.3$, $N \!= \!99$. The time evolution of the chain is shown  (see also movie \ref{EvolutionForKnotting4}). Here $k_f$ denotes the knotted fraction based on the stochastic closure scheme described in the paper.}
\label{EvolutionForKnotting4}
\end{figure}

We have found 13 examples of the unknotting-knotting transition: for fibers with $A=1.4$, initially oriented as $I_0=b$, and for fibers with $A=2.1,2.2,2.3$, initially oriented as $I_0=c$. In most cases, the unknotting-knotting transition is observed more than once during evolution of the same fiber. In figures \ref{EvolutionForKnotting1}-\ref{EvolutionForKnotting4}  we plot the time-dependent 
 knotted fraction $k_f$ based on the stochastic closure scheme,  and display some unknotted and some knotted configurations. Also, in the figures we use different colors to help highlight the complex topology of the complicated chain.

Two examples of the unknotting-knotting transition with $A=2.2$, visible in the time evolution of $k_f$ in figure \ref{EvolutionForKnotting1}, are
shown in movie \ref{EvolutionForKnotting1}. In particular, in figure \ref{EvolutionForKnotting1} it is also illustrated how the chain first unknots as $k_f$ decreases below $0.1$, 
and then knots again, as evidenced by $k_f$ then increasing to approach unity. Eventually, however, the chain is visibly unknotted and $k_f$ remains near zero. The other examples of the unknotting-knotting transitions that we have uncovered are shown in  movies \ref{EvolutionForKnotting2}-\ref{EvolutionForKnotting4} and the plots of the time-dependent $k_f$ in figures \ref{EvolutionForKnotting2}-\ref{EvolutionForKnotting4}. 
Specifically, for $A=1.4$, as displayed in figure \ref{EvolutionForKnotting2} and movie \ref{EvolutionForKnotting2}, a knotted fiber is unknotted after time $t\approx 330$ and knots again at time $t\approx 380$. At the end of the simulation, the chain remains in the knotted configuration, with $k_f$ close to one. 
The evolution of a fiber with $A=2.1$ is shown 
in figure \ref{EvolutionForKnotting3} and movie \ref{EvolutionForKnotting3}. Again, the initially knotted chain unknots and knots again several times.
Finally, the chain is knotted and stays in this form to the end of the simulation. 

In yet another example, the dynamics of a fiber with $A=2.3$ is presented in the figure \ref{EvolutionForKnotting4} and movie \ref{EvolutionForKnotting4}. The fiber 
unties and knots again many times, as shown in the graph of the knotting fraction $k_f$. 
Finally, the fiber unties and stays unknotted until the end of the computation. Thus, we have examples, at least for the time duration of our simulations, of some cases that end up knotted and other cases that end up unknotted. 

All of the configurations displayed in figures \ref{EvolutionForKnotting1}-\ref{EvolutionForKnotting4}, can be seen in three dimensions, and rotated using the corresponding MATLAB  fig-files, given in the Supplementary Data and labeled to match the numbers of the related figures; for example, the file fig5a.fig corresponds to the first frame of figure 5 in the text.

\subsection{Characterizing configurations that display the unknotting-knotting transition}

Comparing the different simulations with each other, e.g. figures \ref{EvolutionForKnotting1}-\ref{EvolutionForKnotting4}, we conclude that
the existence of the unknotting-knotting transition is not correlated with 
a final knotted or unknotted state. 
The question then can be asked if the existence of the unknotting-knotting transition for a trefoil is correlated with some specific mode
 of the flexible fiber dynamics, similar to those observed in the simulations that started from the initial candy-cane configuration, which were shown in figure \ref{shapesgraph}. For example, 
the characteristic features of the two main conformational modes where the fiber primary axis tends to be aligned along the vorticity axes (mode I) or the flow direction (mode II), discussed earlier in connection to figure \ref{shapesgraph}, are summarized in table \ref{2modes}. 

\begin{table}[th] 
\centering
\caption{Characteristics of two main shape modes of flexible fibers in shear flow.}\label{2modes}
\begin{tabular}{ccc}
parameter$\setminus$mode &\hspace{0.5cm} I & \hspace{0.6cm}II\\
\hline
$a,b,c$&\hspace{0.5cm}$a \gg b\gtrsim  c$ &\hspace{0.6cm} $a \gg b \gg c$ \\
$\Theta$&\hspace{0.5cm} $\approx 0$ or $\pi$&\hspace{0.6cm} $\approx \pi/2$\\
$\Phi$&\hspace{0.5cm}increasing&\hspace{0.6cm} $\approx \frac{\pi}{2}$ or $\frac{(2k\!+\!1)\pi}{2}$\\
close pairs&\hspace{0.5cm}many&\hspace{0.6cm}a few
\end{tabular}
\end{table}

\begin{figure}[t]
\includegraphics[width=1.\textwidth]{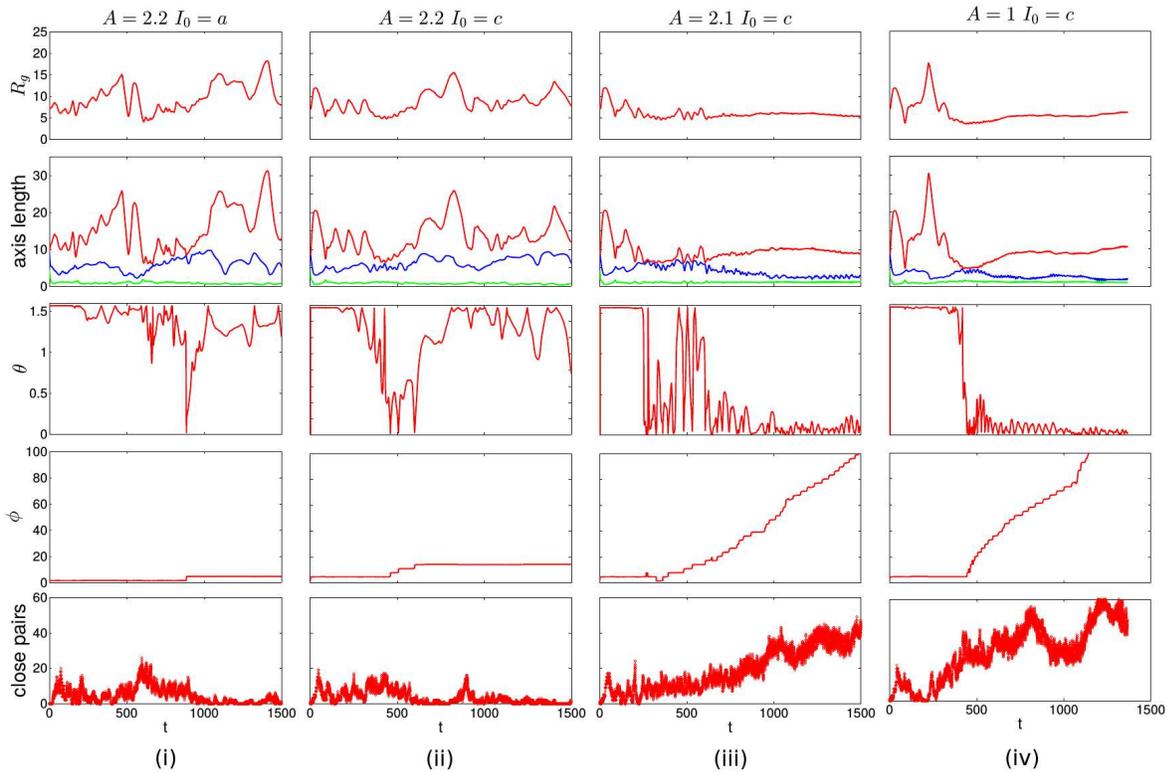}\vspace{-0.6cm}
\caption{Dynamics of trefoils for different values of bending stiffness $A$ and initial orientations $I_0$. The fiber radius of gyration $R_g$,  length of the semi-axes $a,b,c$, orientation $\theta$ and $\phi$ of the primary axes, and the number of close pairs are defined in the main text. The unknotting-knotting transition is observed during evolution shown in columns (ii) and (iii).}
\label{EvolutionForParameters}
\end{figure}

Therefore, in figure \ref{EvolutionForParameters}, we compare the dynamics of trefoils with different bending stiffnesses $A$ and initial orientations $I_0$ in terms of their characteristic modes of conformation. 
We present evolution of the same parameters of the fiber shapes as in figure \ref{shapesgraph}, but this time for fibers that have an initial trefoil shape. 
For two sets of the parameters, corresponding to the dynamics shown in columns (ii) and (iii), the unknotting-knotting transition is observed. For two other sets of the parameters, related to columns (i) and (iv), 
the unknotting-knotting transition is not observed. 

We first compare two cases for which the unknotting-knotting transition is observed, i.e. 
columns (ii) and (iii) in figure \ref{EvolutionForParameters}. For example, the evolution displayed in column (iii) for larger times approximately corresponds to mode I, which is characterized by an elongated shape with relatively large aspect ratios, and the tendency to align with the vorticity direction and roll around it. For comparison, at long times, the dynamics visible in column (ii) approximately corresponds to mode II, which is characterized by a smaller elongation, and no special features of the dynamics related to the vorticity direction. 

On the other hand, comparing columns (i) and (ii) in figure \ref{EvolutionForParameters}, we conclude that the same mode II configuration is reached by fibers for which the unknotting-knotting transition, respectively, did not and did take place. These examples of two topologically different behaviors correspond to different  initial orientations of the trefoil characterized by a single value of $A$. 
The complementary comparison of columns (iii) and (iv) in figure \ref{EvolutionForParameters} demonstrates that fibers 
with and without the unknotting-knotting transition can both later exhibit  the same mode I configuration.  In this case, examples of two topologically different behaviors correspond to the same initial orientation of the trefoil but different values of $A$. 
Therefore, the examples displayed in figure \ref{EvolutionForParameters} illustrate that the existence of the unknotting-knotting transition is not correlated with the final configuration mode I or mode II of the dynamics. 

\begin{figure}[h]
\centering
\includegraphics[width=\textwidth]{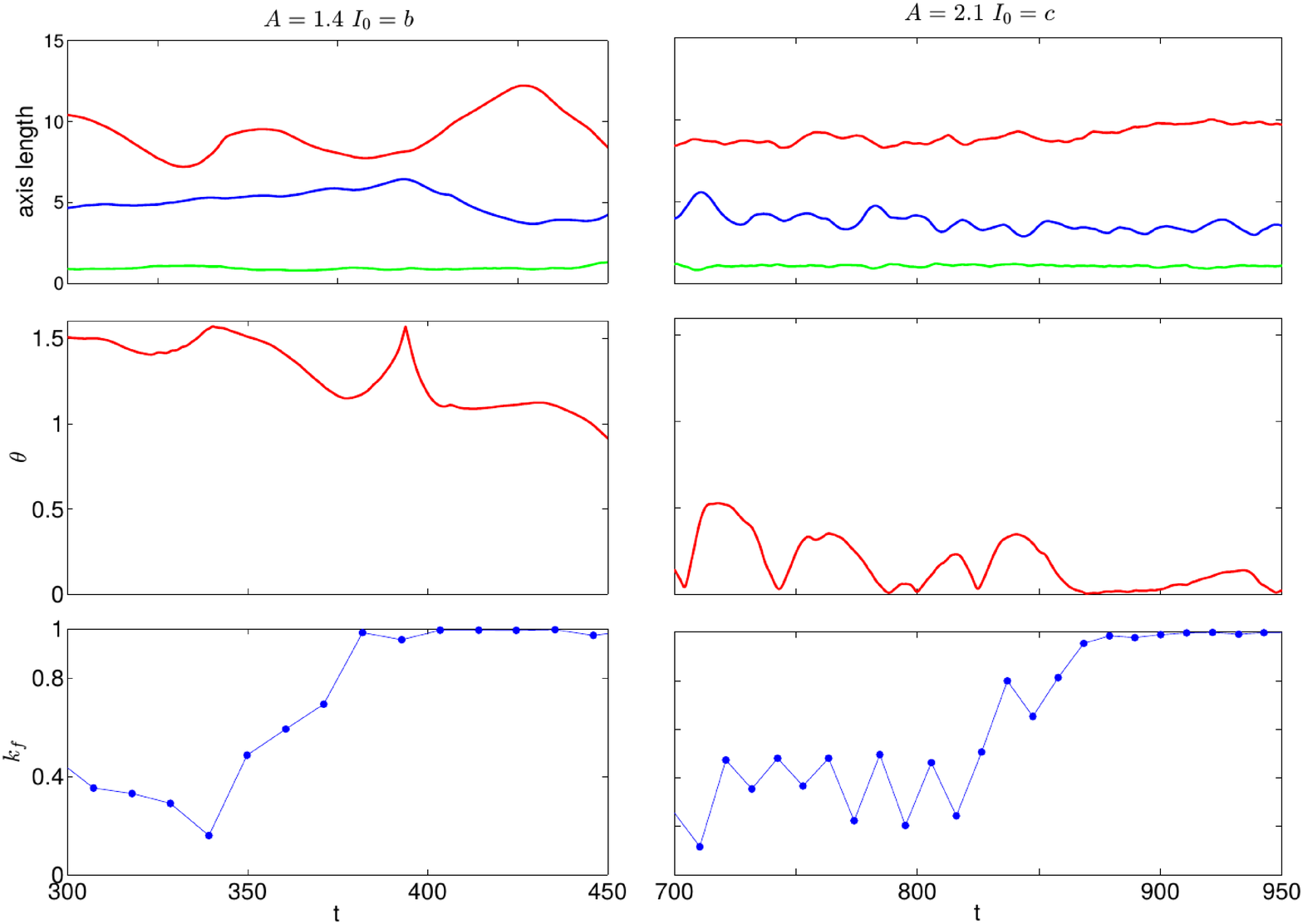}
\caption{The process of creating a knot can take place when the fiber is in mode I (right) or II (left).}\label{noInoII}
\end{figure}

Finally, we investigate if the process of creating a knot is correlated with any simultaneous specific features of the fiber shape.  
Two examples shown in figure~\ref{noInoII} contribute to a negative answer. In both cases, the fiber starts from very small knotted fraction $k_f$, which increases up to almost one. 
However, in the first case (shown in the left part of figure~\ref{noInoII}), the fiber is in mode II, and in the second case (shown in the right part of figure~\ref{noInoII}), 
 the fiber is in mode I. To conclude this discussion, we note that the process of knot creation takes place both for fiber shapes that are compact (figures \ref{EvolutionForKnotting2}, \ref{EvolutionForKnotting3}) and loose (figures \ref{EvolutionForKnotting1}, \ref{EvolutionForKnotting4}).
 
\section{Summary and conclusions}
We have focused on the question of whether  a simple shear flow can tie a knot in a flexible fiber. If yes, then can we predict the future existence of a knot based on the fiber flexibility and its initial shape and orientation? Also, is the knot a transient effect or a stable configuration? To address these questions, the fiber was modeled as a chain of $N$ beads connected by very stiff springs with additional elastic forces acting against the fiber bending. The fiber evolution was determined using the numerical code {\sc hydromultipole}, based on the accurate multipole algorithm for solving the Stokes equations with no-slip boundary conditions on the bead surfaces.
In addition, to characterize the topology a knot on a flexible open fiber was defined as in Refs. [33-34] and associated with the knotted fraction $k_f$ close to unity. 

In the first part of our study, we investigated the evolution of fibers of different flexibilities, but with the same initial orientation and the same untied shape (with a vanishing knotted fraction). We concluded that for such a simple initial configuration, a change of flexibility is not sufficient to cause formation of a knot, although evidence of slipknots was observed.

\begin{figure}[h]
\parbox[c]{7.5cm}{
\includegraphics[width=7.5cm]{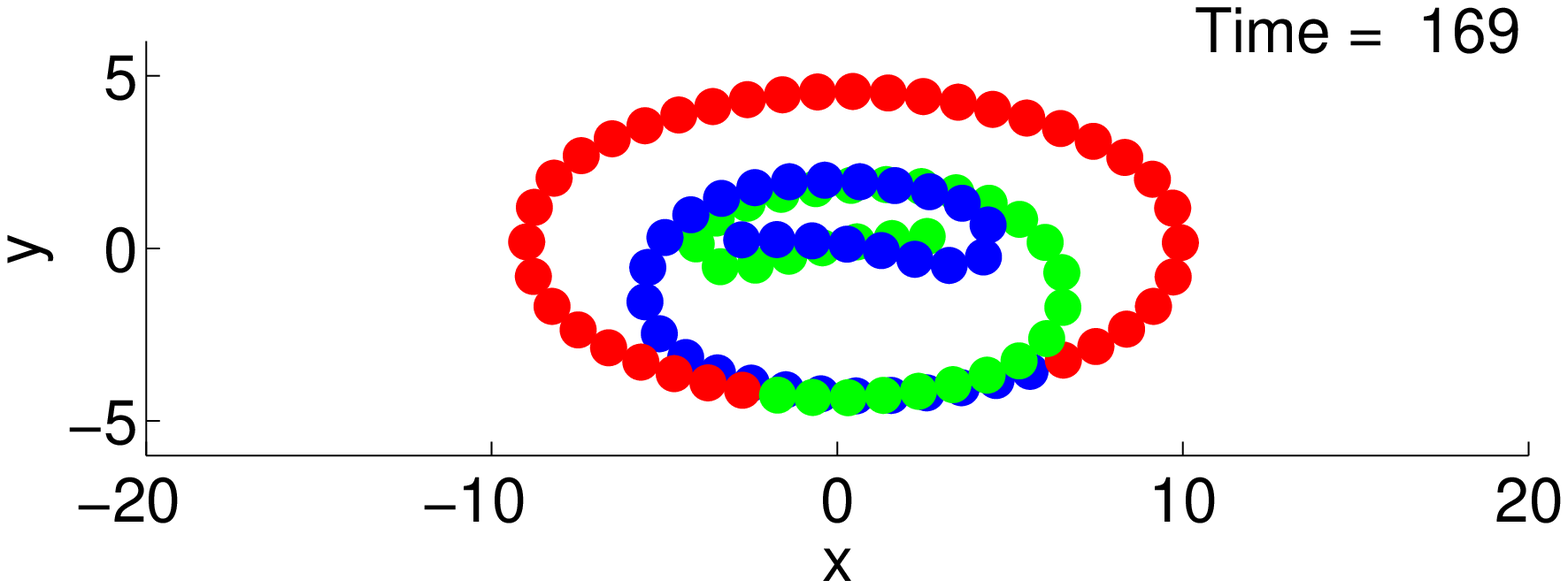}} \parbox[c]{0.8cm}{\LARGE{$\rightarrow$}} \parbox[c]{7.5cm}{\includegraphics[width=7.5cm]{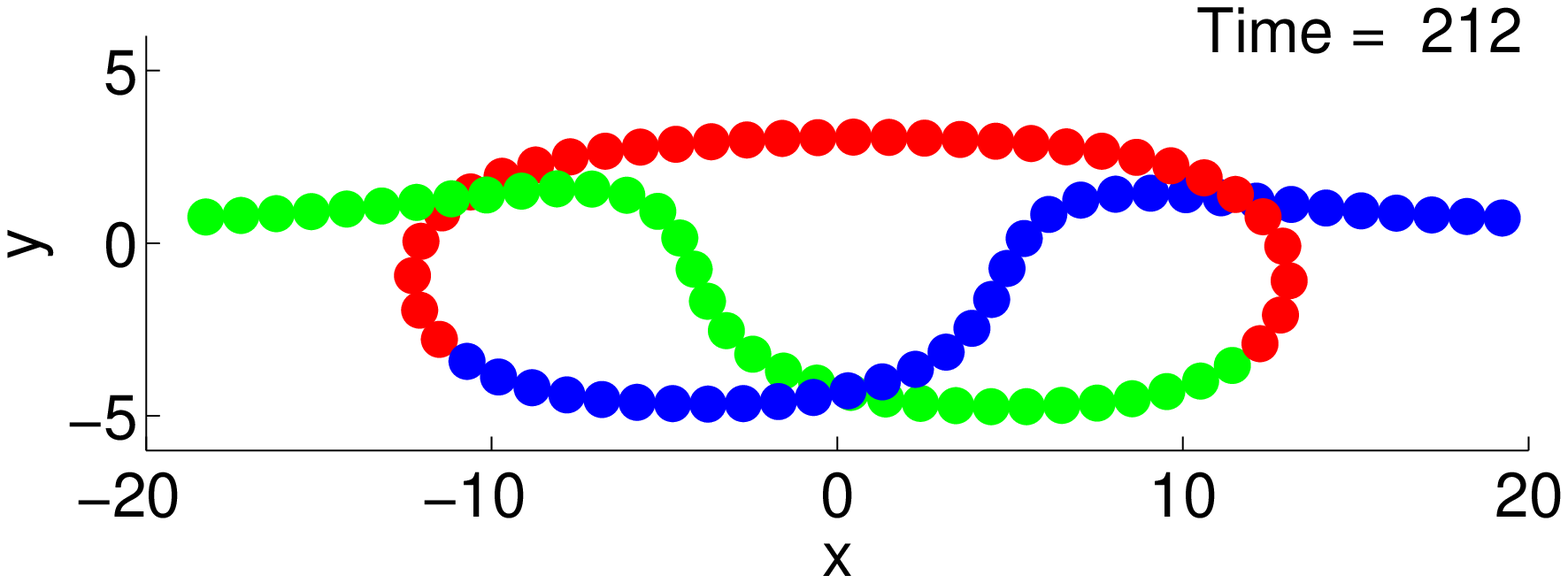}}
\parbox[c]{7.5cm}{\includegraphics[width=7.5cm]{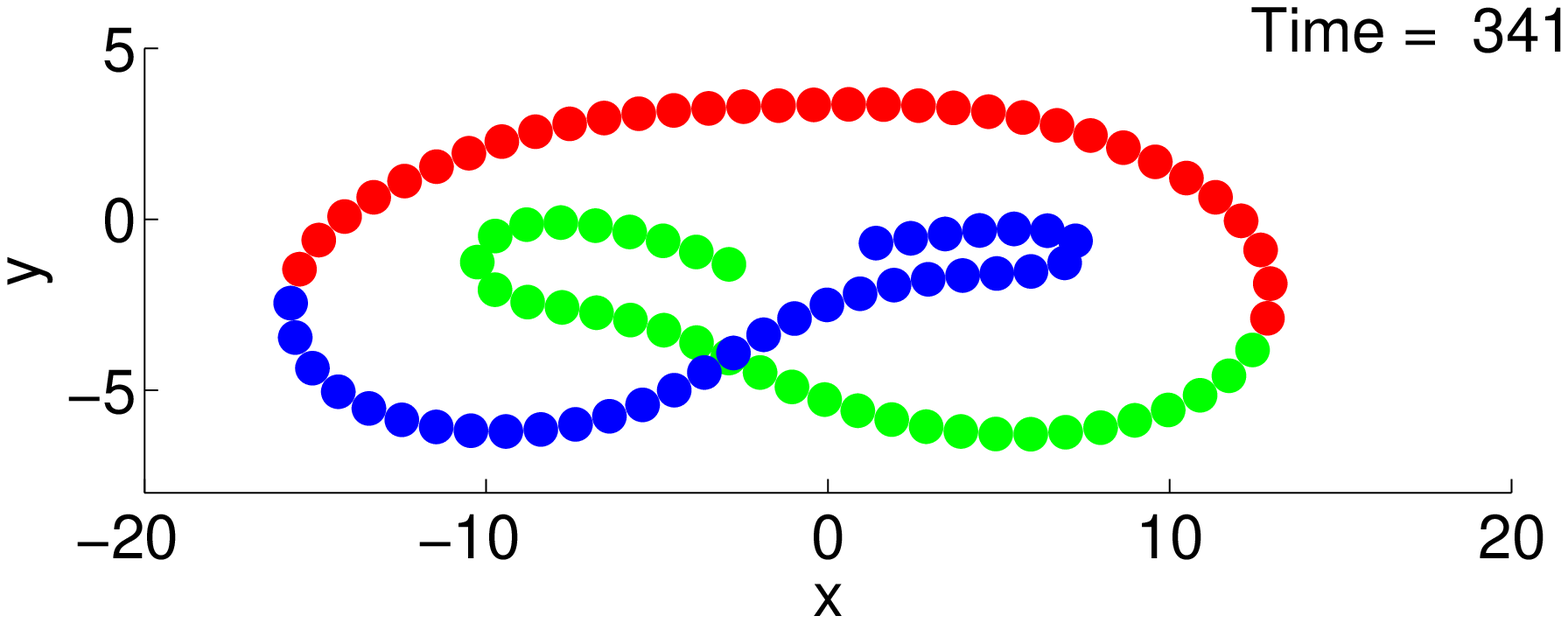}} \parbox[c]{0.8cm}{\LARGE{$\rightarrow$}} \parbox[c]{7.5cm}{\includegraphics[width=7.5cm]{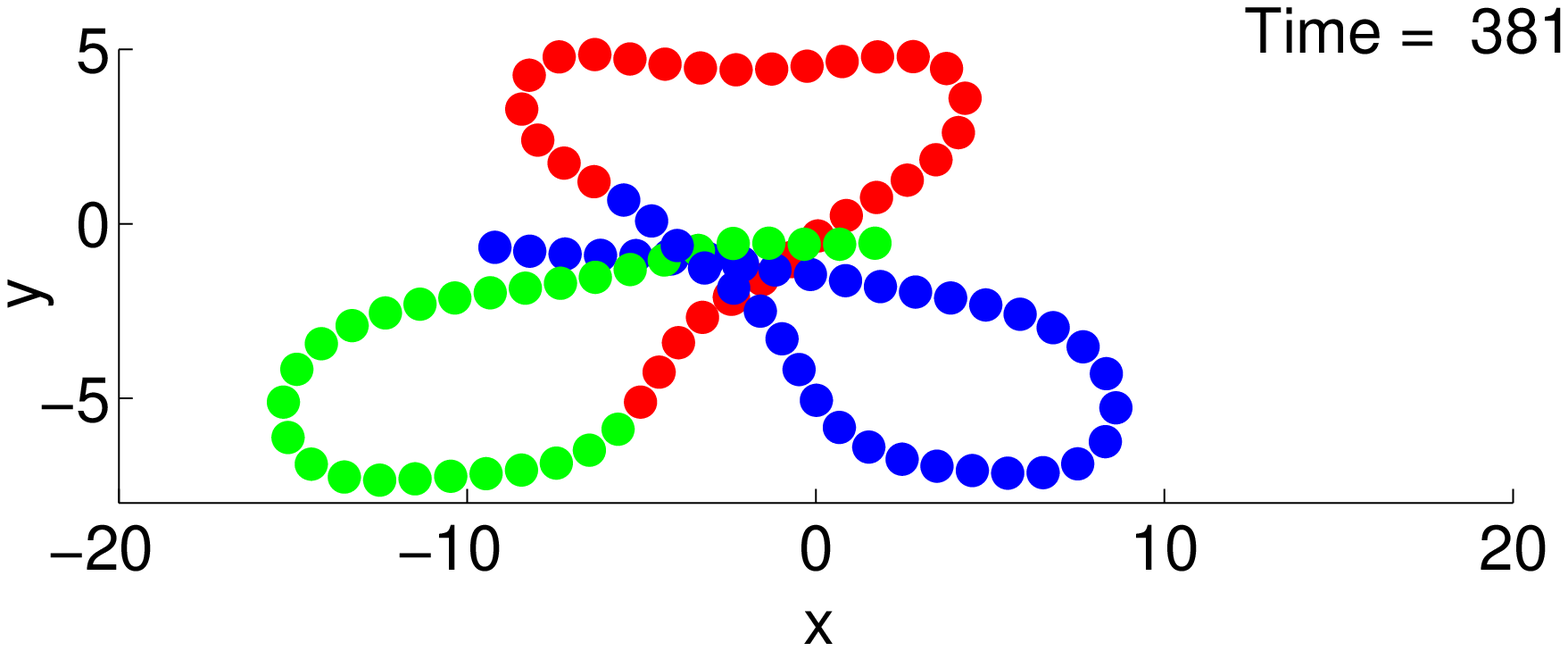}}
\caption{From unknotted to knotted: examples that illustrate that simple shear flow can tie a knot in a flexible fiber. Knotted fraction close to zero (left) is followed by knotted fraction close to one (right). $A=2.3$ and $N=99$.}\label{summary}
\end{figure}

Then, in the second part of our study, we constructed a more complicated initial shape: a trefoil with the knotted fraction equal to one. We used the same initial shape at different initial orientations and for fibers of different flexibilities, and we followed their evolution in a shear flow. A few of the more flexible fibers remained knotted, but most of the fibers untied after some time. In particular, we then observed that some of these untied fibers again became knotted, as illustrated in figure  \ref{summary} for $A=2.3$ and $N=99$.  
In this figure we display two shapes with the knotted fraction close to zero, which later on lead to evident knots (with the knotted fraction close to one). 

Therefore, indeed, a simple shear flow can tie a knot on a flexible fiber. Sometimes, such an unknotting-knotting transition repeated several times during the evolution of the fiber dynamics. We observed 13 such  unknotting-knotting transitions in 4 (out of 29) simulations, for 4 different (moderate) flexibilities and 2 different initial orientations. Based on all of our observations, we conjecture that the formation of knots is a transient effect 
and may possibly be related to the chaotic properties of the dynamics of this $N$ bead system. 

To conclude, we have analyzed basic features of the dynamics of long flexible fibers in unbounded steady shear flow. In particular, we 
give examples of flexible fibers, initially shaped as open trefoils, which are later untied and then knotted again, sometimes several times, by the flow. This finding provides a new perspective for future studies of topological structures of long flexible non-Brownian chains evolving as they are entrained by ambient flows, and possibly changing the macroscopic transport properties.  Finally, as we noted in the introduction, the numerical solution of this problem requires the solution of $3N$ coupled nonlinear ordinary differential equations for the positions of the $N$ beads. We should expect such a dynamical system to have all of the attributes of chaos.
How the observed dynamics of knot formation and unknotting may (or may not) link to the underlying chaotic trajectories is a subject for future investigations.
\\

\ack
HAS thanks the NSF for support from grant CBET-1234500.
AMS, MLEJ and EW were supported in part by the Polish National Science Centre, 
Grant No. 2011/01/B/ST3/05691. MLEJ benefited from scientific activities of the COST Action MP1305.

\appendix

\section{Initial configurations of a ``candy-cane" shape}
\label{AppendixCandyCane}

\begin{figure}[ht]\begin{center}
\includegraphics[width=6cm]{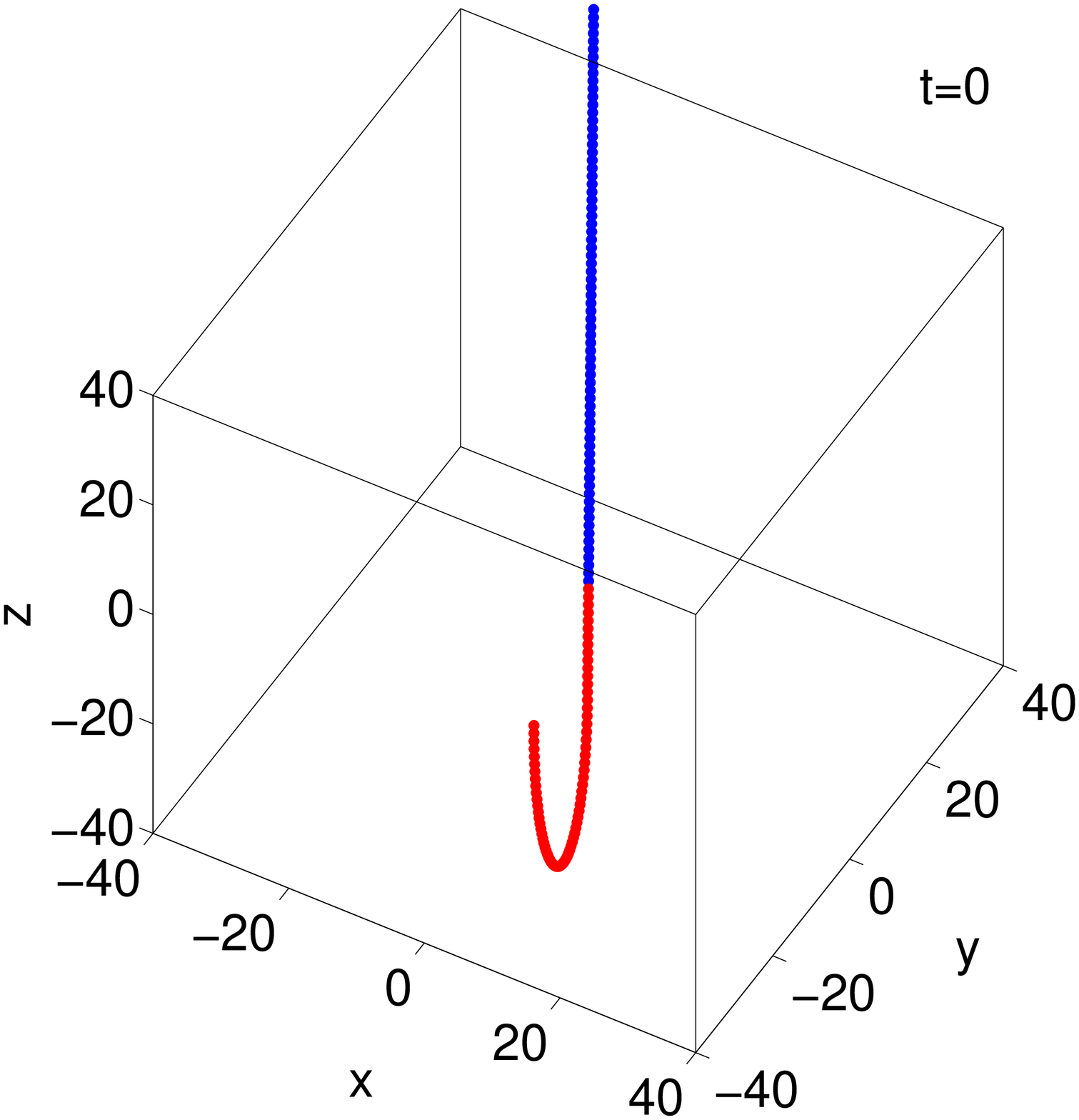}\end{center}
\caption{Initial candy-cane shape.}
\label{cc}
\end{figure}

In the first part of our study, we started simulations from initial configuration of a fiber made of 152 beads, shown in figure~\ref{cc}. 
This `candy-cane' shape  was generated from the equation
$ \left(20 \cos s, 20 \sin s+20, 0 \right) $
with $s$ ranging from $-\pi/2$ to $\pi/2$ and the step $ds = 0.05105$. This equation is a semicircle of radius $20$, and $ds$ has been chosen in such a way that the distance between each point is equal to $\ell_0 = 1.02$ bead diameters. 
A rod starting from $(-91.8,0,0)$, ending at $(0,0,0)$ is then connected; the points are again $\ell_0 = 1.02$ apart, and described by:
\begin{equation}
 \left[ -1.02\left( 91-s \right), 0, 0 \right], \end{equation}
with $s$ ranging from $1$ to $90$ in increments of $1$. 
A rotation matrix is randomly generated and applied to these coordinates to rotate them out of the plane $xy$. In figure~\ref{cc}, we show the initial condition 
 which is used as an input for the simulations displayed in figure 2 and discussed in this paper.

\section{Initial configurations of a trefoil shape}
\label{AppendixTrefoil}

The trefoil shape is constructed with the use of the following equation,
\begin{equation}
 \left[ \sin s + 2 \sin(2s), \cos s - 2 \cos(2s), -\sin(3s)  \right],
\end{equation}
from which 100 evenly spaced points are chosen with $\ell_0=1.02$.
Since the points are evenly spaced along the path of the curve, there are slight variations in the straight-line distance between each point.
One point was then removed to `open' the trefoil as an open curve.
The resulting trefoil with $N = 99$ and $k_f \approx 0.98$ is shown in figure \ref{InitialConditions}a and denoted as $I_0=a$.

\section{Knot Invariants}
\label{KnotInvariantsAppendix}

After our open curve is closed by the stochastic closure scheme, it can be analyzed for its polynomial knot invariant. There are by now many different knot invariants, which have varying ability to distinguish knots; the first one, which was discovered in 1928 by Alexander, is one of the simplest to compute. It has some drawbacks, such as the inability to distinguish handedness (i.e. it cannot distinguish between two mirror images), but as we only need to distinguish between an `unknot' and a `knotted' curve to calculate the knot fraction, the Alexander polynomial is sufficient. 

The invariant is calculated by projecting a curve into 2D, then identifying `crossing points' and `regions', while keeping track of the `overlying' and `underlying' branches at each crossing. If there are $n$ crossings, there will be $n+2$ regions in the knot diagram. Depending on the orientation of the overlying/underlying branches at each crossing, the regions adjacent to the crossing are assigned different indices. A $n \times (n+2)$ incidence matrix is constructed from the indices at each crossing and region, and two columns corresponding to adjacent regions are removed, to form a $n \times n$ matrix. The determinant of the matrix is then the Alexander polynomial. More details, and proof that removing two arbitrary columns (provided they refer to adjacent regions) results in the same normalized Alexander polynomial, can be found in his 1928 paper~\cite{Alexander1928}. 

We have written a custom code to compute the Alexander polynomial of any configuration, given the $(x,y,z)$ coordinates of each bead in the chain, following the algorithm described by Alexander, with the help of functions from other Alexander polynomial calculators~\cite{rytin}. 


\end{document}